\documentclass[traditabstract, longauth, letter]{aa} 

\usepackage{graphicx}
\usepackage{txfonts}
\begin{document}
   \title{Herschel photometric observations of the low metallicity dwarf galaxy NGC 1705}

\authorrunning{O'Halloran et al.}

\titlerunning{Herschel observations of NGC 1705}

   \author{B. O'Halloran\inst{1}\thanks{email:b.ohalloran@imperial.ac.uk},
  	  M. Galametz\inst{2},
	  S. C. Madden\inst{2}, 
	  R. Auld\inst{3},
          M. Baes\inst{4},
          M. J. Barlow\inst{5},
          G. J. Bendo\inst{1},
          J. J. Bock\inst{6},
          A. Boselli\inst{7},
          M. Bradford\inst{6},
          V. Buat\inst{7},
          N. Castro-Rodriguez\inst{8},
	  P. Chanial\inst{1},
          S. Charlot\inst{9},
          L. Ciesla\inst{7},
	  D. L. Clements\inst{1},
	  D. Cormier\inst{2},
          A. Cooray\inst{10},
	  L. Cortese\inst{3},
	  J. I. Davies\inst{3},
          E. Dwek\inst{11},
	  S. A. Eales\inst{3},
	  D. Elbaz\inst{2},
	  F. Galliano\inst{2},
	  W. K. Gear\inst{3},
          J. Glenn\inst{12},
	  H. L. Gomez\inst{3},
	  S. Hony\inst{2},
	  K. G. Isaak\inst{3},
          L. R. Levenson\inst{6},
          N. Lu\inst{13},
          K. Okumura\inst{2},
          S. Oliver\inst{14},	 
          M. J. Page\inst{15},
	  P. Panuzzo\inst{2},
	  A. Papageorgiou\inst{3},
          T. J. Parkin\inst{16},
          I. Perez-Fournon\inst{8},
	  M. Pohlen\inst{3},
          N. Rangwala\inst{12},
           E. E. Rigby\inst{17},
          H. Roussel\inst{9},
	  A. Rykala\inst{3},
          N. Sacchi\inst{18},
	  M. Sauvage\inst{2},
          B. Schulz\inst{13},
          M. R. P. Schirm\inst{16},
	  M. W. L. Smith\inst{3},
          L. Spinoglio\inst{18},
          S. Srinivasan\inst{9},
          J. A. Stevens\inst{19},
          M. Symeonidis\inst{15},
	  M. Trichas\inst{1},
          M. Vaccari\inst{20},
          L. Vigroux\inst{9},
          C. D. Wilson\inst{16},
          H. Wozniak\inst{21},
          G. S. Wright\inst{22},
	  W. W. Zeilinger\inst{23}
          }

  \institute{Astrophysics Group, Imperial College, Blackett Laboratory, Prince Consort Road, London SW7 2AZ, UK 
           \and
		Laboratoire AIM, CEA/DSM - CNRS - Universit\'e Paris Diderot, Irfu/Service d'Astrophysique, 91191 Gif sur Yvette, France
	 \and
		School of Physics and Astronomy, Cardiff University, Queens Buildings The Parade, Cardiff CF24 3AA, UK
	\and	
		Sterrenkundig Observatorium, Universiteit Gent, Krijgslaan 281 S9, B-9000 Gent, Belgium
	 \and
		Department of Physics and Astronomy, University College London, Gower Street, London WC1E 6BT, UK	 
	 \and
		Jet Propulsion Laboratory, Pasadena, CA 91109, United States; Department of Astronomy, California Institute of Technology, Pasadena,  CA 91125, USA
	\and
		Laboratoire d'Astrophysique de Marseille, UMR6110 CNRS, 38 rue F. Joliot-Curie, F-13388 Marseille France
	\and
		Instituto de Astrofisica de Canarias, C/ Vía L\'actea, s/n, E38205 - La Laguna (Tenerife). Spain s/n, E-38200 La Laguna, Spain	
	\and
		Institut d'Astrophysique de Paris, UMR7095 CNRS, Universit\'e Pierre \& Marie Curie, 98 bis Boulevard Arago, F-75014 Paris, France	
	\and
		Department of Physics \& Astronomy, University of California, Irvine, CA 92697, USA
	\and
		Observational  Cosmology Lab, Code 665, NASA Goddard Space Flight Center Greenbelt, MD 20771, USA	
	\and
		Department of Astrophysical and Planetary Sciences, CASA CB-389,  University of Colorado, Boulder, CO 80309, USA	
	\and
		Infrared Processing and Analysis Center, California Institute of Technology, Mail Code 100-22, 770 South Wilson Av, Pasadena, CA 91125,  USA	
	\and
		Astronomy Centre, Department of Physics and Astronomy, University of Sussex, UK	
	\and
		Mullard Space Science Laboratory, University College London, Holmbury St Mary, Dorking, Surrey RH5 6NT, UK	
	\and
		Dept. of Physics and Astronomy, McMaster University, Hamilton, Ontario, L8S 4M1, Canada	
	 \and
		School of Physics \& Astronomy, University of Nottingham, University Park, Nottingham NG7 2RD, UK
	\and
		Istituto di Fisica dello Spazio Interplanetario, INAF, Via del Fosso del Cavaliere 100, I-00133 Roma, Italy	
	\and
		Centre for Astrophysics Research, Science and Technology Research Centre, University of Hertfordshire, College Lane, Herts AL10 9AB, UK
	\and
		University of Padova, Department of Astronomy, Vicolo Osservatorio 3, I-35122 Padova, Italy
        \and
		Observatoire Astronomique de Strasbourg, UMR 7550 Universit\'e de  Strasbourg - CNRS, 11, rue de l'Universit\'e, F-67000 Strasbourg       
       \and
                UK Astronomy Technology Center, Royal Observatory Edinburgh, Edinburgh, EH9 3HJ, UK
       \and
		Institut f\"ur Astronomie, Universit\"at Wien, T\"arkenschanzstr. 17,  A-1180 Wien, Austria
}

   \date{Submitted to A\&A Herschel Special Issue}


\abstract{We present Herschel SPIRE and PACS photometeric observations of the low metallicity (Z $\sim$ 0.35Z$_{\odot}$) nearby dwarf galaxy, NGC 1705, in six wavelength bands as part of the Dwarf Galaxy Survey guaranteed time Herschel Key Program. We confirm the presence of two dominant circumnuclear IR-bright regions surrounding the central super star cluster that had been previously noted at mid-IR wavelengths and in the sub-mm by LABOCA. On constructing a global spectral energy distribution using the SPIRE and PACS photometry, in conjunction with archival IR measurements, we note the presence of an excess at sub-mm wavelengths. This excess suggests the presence of a signiÞcant cold dust component within NGC 1705 and was modeled as an additional cold component in the SED.
Although alternative explanations for the sub-mm excess beyond 350 $\mu$m, such as changes to the dust emissivity cannot be ruled out, the most likely explanation for the observed submillimetre excess is that of an additional cold dust component. }



   \maketitle

%

\section{Introduction}

Understanding the origin and evolution of dwarf galaxies is important in modern observational cosmology. Current models for galaxy formation fall into two distinct categories: construction from the ``top-down'' or from the ``bottom-up''. Which scenario is correct? Unfortunately, the answer is still somewhat ambiguous and left open to interpretation as both models face critical problems, in particular due to a lack of detailed, high resolution studies at redshifts corresponding to the peak era of galaxy formation. Bottom-up models over-predict the number and the mass spectrum of satellites seen around galaxies such as our own and there are inconsistencies with the timescale of the build up of larger galaxies (e.g., Prantzos \& Silk 1998). A major source of this ambiguity may arise from the fact that accurate measurements of the interstellar medium (ISM) properties and star formation histories of dwarf galaxies have been lacking for the most part.  To address this crucial issue, we have begun to focus our study of galaxy evolution onto analogues of these earliest galaxy building blocks right on our door step - nearby, low metallicity dwarf galaxies - where we can begin to discern the evolutionary processes at work in these objects at very high spatial resolution.  Such objects are  the most common type of galaxy in the current epoch, making up 95\% of the Local Group galaxies alone (Mateo 1998).
\begin{figure*}
\centering
\includegraphics*[angle=270, width=15cm]{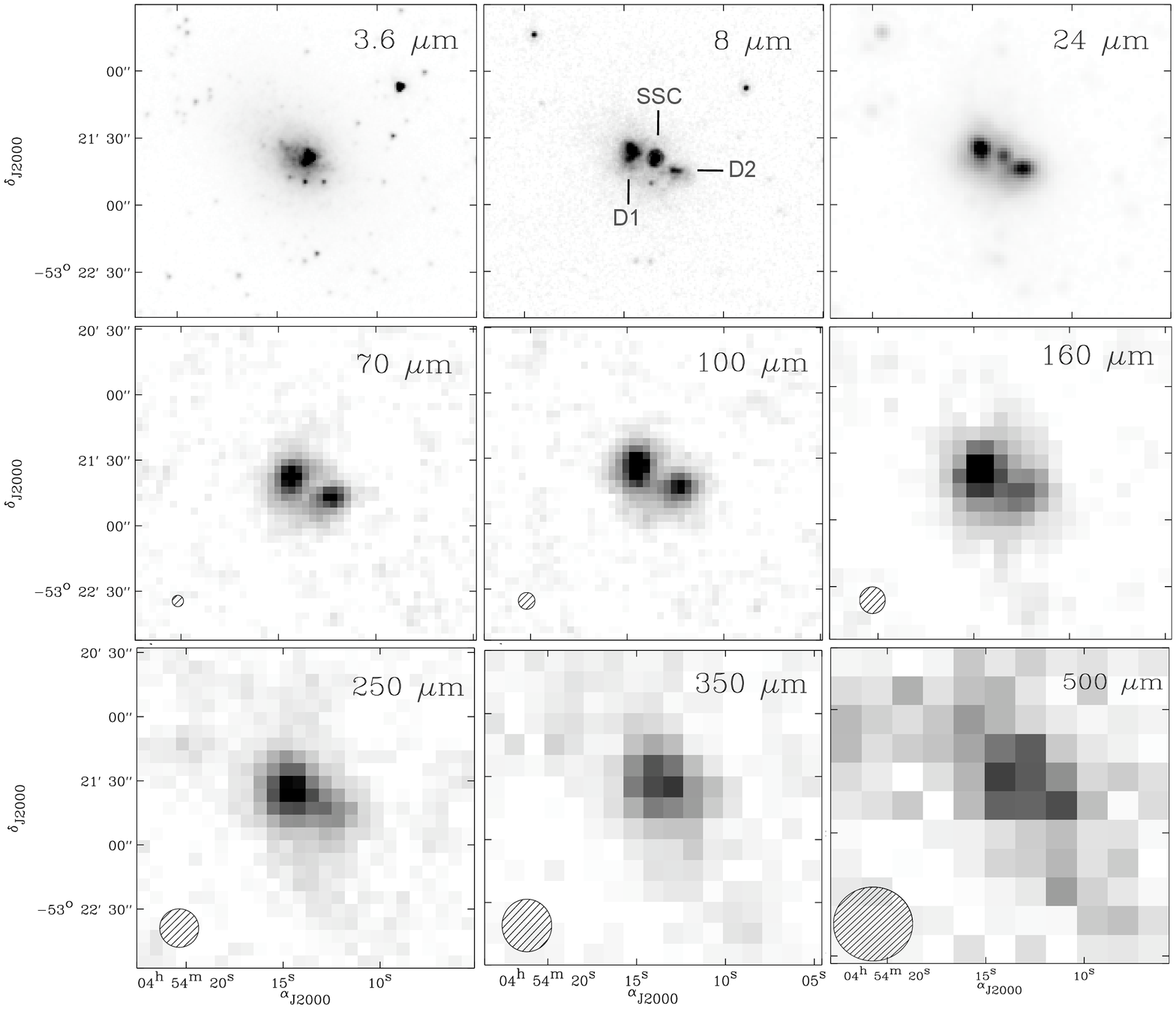}
\caption{IRAC/MIPS (top) PACS (middle) and SPIRE (bottom) maps of NGC 1705. The top three panels are the Spitzer IRAC and MIPS maps (l-r) at 3.6, 8 and 24 $\mu$m, the middle panels are the PACS maps (l-r)  at 70, 100 and 160 $\mu$m, while the bottom three panels are the maps at 250, 350 and 500 $\mu$m respectively. The central SSC and the eastern (D1) and western IR components are noted on the IRAC 8 $\mu$m map, whilst the beam FWHM values for the PACS and SPIRE maps are also illustrated.}
\label{photo_plots}
\end{figure*}

As part of this effort to understand the nature of the evolution of dwarf galaxies, we present Herschel{\footnote{Herschel is an ESA space observatory with science instruments provided by European-led Principal Investigator consortia and with important participation from NASA.}} (Pilbratt 2010) photometric observation of the nearby (5.1 $\pm$ 0.6 Mpc (Tosi et al. 2001)) dwarf starburst galaxy, NGC 1705. The galaxy is dominated optically by a massive central super star cluster (SSC) NGC 1705-1 (Meurer et al. 1995), whilst studies in the mid and far-IR (Cannon et al. 2006; Galametz et al. 2009) reveal the presence of two bright infrared regions flanking the central SSC, offset by $\sim$250 pc from the SSC, with these off-nuclear regions dominating the global IR emission of NGC 1705. This galaxy provides an ideal environment for exploring the effects of ongoing, massive star formation on the environment within a dwarf galaxy (Cannon et al. 2006), given its sub-solar nebular metallicity (Z $\sim$35\% Z$_{\odot}$;  (Lee et al. 2004) and large reservoir of gas (Meurer et al. 1998), as we can trace the effects of the SSC on the surrounding interstellar medium, and in particular, can characterize the nature of dust and PAH emission. 

\section{Observations}

NGC 1705 was observed as part of the Dwarf Galaxy Survey programme (PI. S. Madden), a Guaranteed Time (GT) key program with the objective of mapping the dust and gas in 51 nearby dwarf galaxies, sampling a broad metallicity range of 1/50 to 1/3 Z$_{\odot}$. 

\subsection{SPIRE observations and data processing}

The galaxy was observed by SPIRE (Griffin et al. 2010) at 250, 350 and 500 $\mu$m for a total of 733 seconds. in scan-map mode with scanning rate 30"/sec, with the final map covering roughly 16 x 16 arcmin. The measured 1 $\sigma$ noise level are 5, 6 and 7 mJy beam$^{-1}$ at 250, 350 and 500 $\mu$m respectively; the noise levels in the images are dominated by confusion. The data were processed using the HIPE pipeline (see Pohlen et al. (2010) for a detailed description, Swinyard et al. (2010) for calibration accuracy and Bendo et al. (2010b) for details on the destriper). The pipeline produces maps with a pixel size of 6.0, 10.0 and 14.0" at 250, 350 and 500 $\mu$m respectively.  The ICC has released some interim small correction factors to improve the preliminary calibration. All flux values derived using the current standard calibration file for the flux conversion, are multiplied by 1.02, 1.05, and 0.94, for the 250$\mu$m, 350$\mu$m, and 500$\mu$m maps, giving final global fluxes of 0.85 $\pm$ 0.13, 0.38 $\pm$ 0.06 and 0.26 $\pm$ 0.04, respectively, using an aperture of 72" as per Galametz et al. (2009). The uncertainty in the flux calibration is of the order of 15\%.

\subsection{PACS data processing}

The galaxy was observed by PACS (Poglitcsh et al. 2010) at 70, 100 and 160 $\mu$m in scan map mode for a total of 1504 seconds, with the final map covering roughly 7 x 6 arcmin. The PACS data were reduced starting from the Level 0 product using the HIPE version 3.0 data reduction software and the HIPE responsivity calibration file version 1, corrected by factors provided by the ICC. We applied the basic steps of the standard reduction pipeline to perform the data reduction to the Level 1, where we mask the bad and saturated pixels, apply a flat field correction and perform a multi-resolution median transform (MMT) deglitching and apply a second order deglitching procedure to the data. We mask the data contributing to the bright structures in the data cube and perform polynomial fits on half scan legs to subtract the baselines. This process removes most of the drifts of the maps.  Two dimensional maps are finally constructed using the MadMap procedure of HIPE. Striping is visible in a single scan but is mostly removed by the cross scan. The pipeline produces maps with a pixel size of 3.2, 3.2 and 6.4" at 70, 100 and 160 $\mu$m respectively. In this version of the responsivity calibration file, an error still exists in the flux scale. The final reduced global fluxes have to be scaled down by 1.05, 1.09 and 1.29 for the blue (PACS 70), green (PACS 100) and red (PACS 160) bands, giving 1.05  $\pm$ 0.12, 1.22  $\pm$ 0.13 and 1.18  $\pm$ 0.12 Jy, respectively, using an aperture of 72" as per Galametz et al. (2009). The uncertainty in the flux calibration is of the order of 10\%.

\section{Results \& Discussion}

\subsection{Morphology of NGC 1705}

The reduced SPIRE and PACS maps of NGC 1705 are presented in Figure 1. To perform aperture photometry on individual sources with NGC 1705 for both the SPIRE and PACS images, we used the IRAF package $\it{apphot}$, in which the background is estimated using annuli just outside the boundaries of our galaxy. To determine the flux densities at MIPS bands, as a comparison to the PACS fluxes, the observations are convolved and re-gridded to a common resolution (FWHM MIPS 160 $\mu$m: 40"). We use convolution kernels (Gordon et al. 2008; Bendo et al. 2010) which convert a higher resolution PACS point-spread function (PSF) to lower resolution MIPS PSFs using Fourier transforms.

\begin{figure}

\includegraphics[width=9cm]{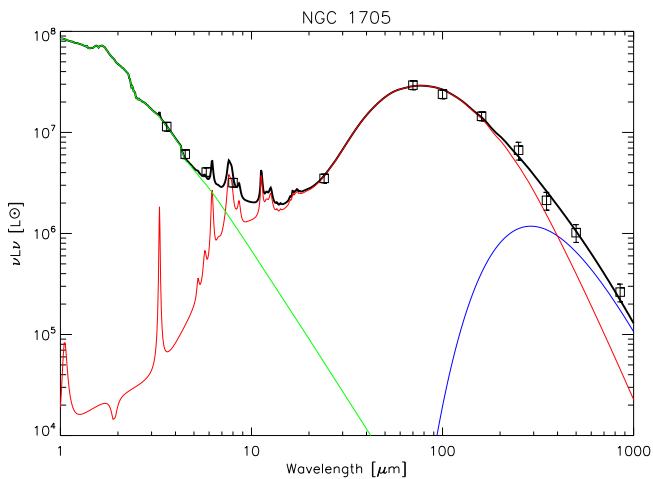}
\caption{Global spectral energy distribution for NGC 1705, with a final cold dust temperature of 5.8 K and $\beta$ =1. The grey curve denotes the combined dust model, the green denotes the stellar component, the red denotes the warm dust/PAH components and the blue denotes the cold dust component. The symbols refer to the Herschel and Spitzer fluxes used to fit the models, in conjunction with the 850 $\mu$m LABOCA flux.}
\label{sed_fit1}
\end{figure}

As can be seen in both sets of images, the morphology of NGC 1705 is quite complex - the far-IR emission is dominated by the off-center sources previously noted by Cannon et al. (2006) and Galametz et al. (2009). As might be expected, the two IR regions are clearly resolved at the shortest PACS wavelengths, with no emission detected from the central SSC which indicates very little in the way of significant dust emission within the central SSC itself. As we move into the SPIRE bands (beyond 250 $\mu$m), the individual components are no longer resolved.  As noted in Galametz et al. (2009), the off-center bright IR fluxes do not have stellar counterparts. There is no indication of the presence of the offset peak to the west of the galaxy in the PACS or SPIRE that had been previously noted by Galametz et al. (2009), nor are there counterparts to the faint 24 $\mu$m sources also noted.  Of note is the fact that the easternmost IR source (D1, in the nomenclature of Cannon et al. (2006)) is the brightest across the different resolved wavelength maps, and is of the order of 1.15-1.45 times brighter than the western source, with the flux ratio peaking at 100 $\mu$m - previous IRS spectroscopy by Cannon et al. (2006) confirms the strongest PAH emission in this region, which in conjunction with the Herschel data is suggestive of a greater concentration/contribution to the dust mass within this region versus both the westernmost region and the SSC.

\subsection{Spectral energy distribution of NGC 1705}

\subsubsection{Cold dust component scenario}

To quantify the contributions to the IR luminosity from various dust populations within NGC 1705, we attempted to fit modified blackbody functions to the SPIRE and PACS measurements in an effort to obtain dust component luminosities, masses and dust temperatures. On fitting, it became quickly apparent that fitting a single blackbody to the SPIRE and PACS fluxes was not possible, as the SED diverged from a single temperature fit longward of approximately 350 $\mu$m with the 350 and 500 $\mu$m fluxes higher than the fit, a trend that was confirmed with the addition of the LABOCA 850 $\mu$m flux from Galametz et al. (2009). Even the addition of a second modified blackbody function to the fit was unable to fit to a reasonable degree the 350 and 500 $\mu$m fluxes in conjunction with the LABOCA flux  - we tried to fit a number of modified blackbodies with a range of realistic emissivity values between 1 and 2, and were unable to obtain a satisfactory fit that suitably fit the three longest wavelength fluxes.

In light of this, we instead constructed a global spectral energy distribution for NGC 1705, which in addition to the SPIRE and PACS fluxes includes global fluxes from LABOCA and the $\it{Spitzer}$ IRAC and MIPS fluxes from Cannon et al. (2006) to better constrain the mid-IR contribution to the SED, and this global SED is presented in Figure 2. We used a realistic dust radiation model as far as dust grain properties were concerned, a simpler version of the Galliano et al. (2008) model. The sources of emission are the dust grains (molecules like PAHs, silicates, graphites) and old stars, while the dust composition and size distribution adopted were those of Zubko et al (2004), with abundances assumed to be solar. The dust optical properties were taken from Draine and Li (2007) for PAHs, Laor \& Draine (1993) for graphites and Weingartner \& Draine (2001) for silicates. The interstellar radiation field spectral shape was assumed to be that of the Galactic diffuse ISM (Mathis et al 1983). The Dale et al. (2001) prescription was used to link the dust mass to the integrated radiation density it was exposed to - see Galametz et al (2009) for a more complete description of the free parameters of the modelling, among those the total dust mass (M$_{dust}$) and the PAH-to-dust mass (f$_{PAH}$) nomalized to the Galactic value. To account for the observed excess at submm wavelengths, an additional, independent, modified blackbody was added to the model.

We then proceeded on two fronts in terms of fitting the model to the fluxes - a) letting $\beta$ vary and keeping the temperature fixed and b) keeping $\beta$ fixed, and letting the temperature vary. However, we found that when $\beta$ was allowed to vary with a temperature of the cold dust fixed (10K), the SED fit had a tendancy to find beta values less than 0.5. This unrealistic value of $\beta$ (normally with values between 1 and 2) (Bolatto et al., 2000) probably reflects instead, a distribution of dust temperatures. As a result of this, we decided to fix the value of $\beta$ to 1 and let the temperature of the cold dust component vary instead. 

The addition of such a dust component (with a temperature of $\leq$ 10 K) within the global SED of NGC 1705 has been tentatively suggested previously by Galametz et al. (2009), on the basis of the addition of the 850 $\mu$m LABOCA point, with this additional dust  component containing up to 70\% of the total dust content of the system. There have been significant claims of prior detections of very cold dust components within dwarf galaxies, (Galliano et al. (2003, 2005) and Galametz et al. (2009)), but the lack of far-IR and sub-mm fluxes to constrain SEDs sufficiently well has left this open to debate.  If  present, very cold dust can only be found deep in the interior of clouds; it must be protected from stellar light and shielded by an optical depth well above 10 mag. The PACS and SPIRE images provide us with an additional constraint regarding the cold dust, namely it must be located where we see the emission, e.g. the eastern condensation, which strongly is suggestive of high extinction in those locations. With the addition of the three SPIRE photometry points, in particular the 350 and 500 $\mu$m points, the claim for an additional dust component is strengthened. 

Our best-fit modeling is consistent with the need for an additional cold dust component, and leads to a cold dust mass of 4.0 x 10$^{5}$ M$_{\odot}$ with a temperature of the cold dust of 5.8 K (with $\beta$ = 1). For the warmer dust component, we estimate a dust mass of 2.1 x 10$^{3}$ M$_{\odot}$, giving a total dust mass of 4.2 x 10$^{5}$ M$_{\odot}$. The plotted fit in Figure 2 is indeed the best available fit - we attempted to fit a range of dust temperatures to satisfy a reasonable fit to the submillimeter excess. However, it was not until the dust temperatures reached less than 10K that we began to obtain acceptable fits - we considered a cold dust component fit with a temperature of 10K and a reduced chi squared value of 1.1 to be the lower limit in terms of an acceptable fit. In terms of obtaining a reasonable lower limit for the dust mass, fitting a cold dust component of 10 K leads to an acceptable reduced chi squared value of 1.1, giving us a cold dust mass of  1.1 x 10$^{5}$ M$_{\odot}$ and a total dust mass of  1.8 x 10$^{5}$ M$_{\odot}$.

The multiband Herschel observations have allowed us to constrain significantly the SED compared to the previous work of  Galametz et al. (2009), where the only sub-mm constraint was the LABOCA flux, allowing us to refine the dust component masses and by extension, obtain new values for the PAH/dust and dust to gas ratios. For the M$_{PAH}$/M$_{dust}$, we get  $\sim$ 2.3 x 10$^{-4}$ - much lower than the Galactic value, and consistent with the PAH emission deficit at very low metallicity (Madden 2002, Galliano et al. 2003, O'Halloran et al. 2006, 2008).  Correspondingly for the dust to gas ratio, we get $\sim$ 1.0 x 10$^{-2}$, using the HI mass from Galametz et al. (2009). The derived dust/gas ratio is rather lower than expected for a low metallicity dwarf galaxy - we could simply be predicting too much dust within our SED modeling, as a result of the use of graphite grains. Serra D\'iaz-Cano \& Jones (2008) warn against the use of graphites in construction of SED modeling, suggesting instead that amorphous carbons could be more relevant to represent the interstellar carbon dust, since they possess a higher emissivity at submm wavelengths and thus require less mass to account for the same emission. Substituting in amorphous carbons to replace the graphite grains in our model (using values from Rouleau \& Martin (1991)) and using the same parameters of $\beta$ and T as before, we obtain reduced warmer (2.86 x 10$^{4}$ M$_{\odot}$) and cold dust masses (9.8 x 10$^{4}$ M$_{\odot}$), with a total dust mass of 1.2 x 10$^{5}$ M$_{\odot}$. These reduced dust masses have the knock-on effect of a higher dust-to-gas ratio ($\sim$ 2.3 x 10$^{-3}$), a value that is much closer to that expected for such a low metallicity galaxy. Both D/G values (from using graphites and amorphous carbons) are still at the lower end of the chemical evolution model used in Galliano, Dwek \& Chanial (2008) - nonetheless, these are values which seem consistent with an object that while possessing a low metallicity, does contain significant amounts of warm and cold dust.

 

\subsubsection{Alternatives to the cold dust component scenario}

We must also consider the possibility that the sub-mm excess detected in NGC 1705 is not as a result of an extra dust component. One hypothesis arguing against such a conclusion is that the sub-mm excess could originate instead from hot ($\sim$100 K) dust with a dust emissivity index =1 and the temperature fluctuations of very small grains (Lisenfeld et al. 2002).  

The excess could also be explained through a change in the emissivity of the cold dust grains. For an example of how the emissivity may change as we move into differing temperature regimes, Dupac et al. (2003) suggest that $\beta$ decreases with increasing temperatures, from $\sim$2 in cold (T $\sim$ 11-20 K) regions to 0.8 to 1.6 in warmer regions (T $\sim$ 35 - 80 K).  In contrast, Paradis et al. (2009) showed that the spectral shape of emissivity are always steeper in the FIR ( $\rm\lambda < 600~{\rm\mu m} $) and flatten in the submm and mm regions. In regions where dust is significantly colder in the molecular phase than in the atomic phase, an increase in the emissivity by a factor of $\rm\simeq $3 was detected only in the FIR, whilst the emissivity for the dust in the atomic and molecular phases become comparable again in the submm and mm wavelength range. The observed break in the emissivity spectrum is in qualitative agreement with the dust emission model of  M\'eny et al. (2007), which invokes quantum effects in amorphous solids to explain the flatness of the observed submm emission spectrum and also produces a break in the emissivity slope around 600  $~{\rm\mu m} $. 

However, one must exercise caution in adopting such an interpretation.  Flux uncertainties, especially in the Rayleigh-Jeans regime, can affect the results for the SED fits as far as temperature and emissivity are concerned, as fitting fluxes near the SED peak produces inaccurate temperature and dust spectral index estimates.  In addition, line-of-sight temperature (and density) variations can also affect the SED fitting (Shetty et al. (2009a,b)). Longer wavelength fluxes in the Rayleigh-Jeans part of the spectrum ($\geq$ 600 $\mu$m) may more accurately recover the spectral index, but both methods are very sensitive to noise (Shetty et al. (2009a,b)). 

{An additional alternative to a cold dust component is spinning dust where the rotational dynamics of very small interstellar grains can explain the 10 - 100 GHz component of the diffuse Galactic background via electric dipole rotational emission under normal interstellar conditions (Draine \& Lazarian 1998a,b). However, Jones (2009) notes that observations by Dickinson et al. (2006), which searched for a microwave emission excess in an H II region, puts an upper limit on the dust emission at 31 GHz and appears to be inconsistent with the spinning dust model.  Given the unsatisfactory evidence arguing against the use of amorphous carbon to explain the sub-mm excess, we conclude that the most likely scenario to explain the observed excess is that of an additional cold dust component.

\begin{acknowledgements}
SPIRE has been developed by a consortium of institutes led by Cardiff University (UK) and including Univ. Lethbridge (Canada); NAOC (China); CEA, OAMP (France); IFSI, Univ. Padua (Italy); IAC (Spain); 
Stockholm Observatory (Sweden); Imperial College London, RAL, UCL-MSSL, UKATC, Univ. Sussex (UK); and Caltech/JPL, IPAC, Univ. Colorado (USA). This development has been supported by
national funding agencies: CSA (Canada); NAOC (China); CEA, CNES, CNRS (France); ASI (Italy); MCINN (Spain); Stockholm Observatory (Sweden); STFC (UK); and NASA (USA).

PACS has been developed by a consortium of institutes led by MPE (Germany) and including UVIE (Austria); KUL, CSL, IMEC (Belgium); CEA, OAMP (France); MPIA (Germany); IFSI, OAP/AOT, OAA/CAISMI, LENS, SISSA (Italy); IAC (Spain). This development has been supported by the funding agencies BMVIT (Austria), ESA-PRODEX (Belgium), CEA/CNES (France),
DLR (Germany), ASI (Italy), and CICT/MCT (Spain).
     
\end{acknowledgements}


\begin{thebibliography}{}

\bibitem[Bendo et al. 2010a]{bendo2010a}	Bendo, G. J.; Wilson, C. D.; Warren, B. E et al., 2010, \mnras 402, 1409B

\bibitem[Bendo et al. 2010b]{bendo2010b} Bendo G. et al., 2010, \aap, this issue 

\bibitem[Bolatto et al. 2000]{bolatto2000} Bolatto, A.D., Jackson, J.M., Wilson, C.D.; Moriarty-Schieven, G., 2000, \apj 532, 909B

\bibitem[Cannon et al. 2006]{cannon2006} Cannon, J.M., Smith, J.D.T., Walter, F., 2006, \apj  647, 293C 

\bibitem[Dale et al. 2001]{dale2001}  Dale, D.A., Helou, G., Contursi, A. et al., 2001, \apj 549, 215D

\bibitem[Dickinson et al. 2006]{dickinson006} Dickinson, C., Cassus, S., Pineda, J. L., et al. 2006, \apj 643, L111

\bibitem[Draine \& Li 2007]{draine2007} Draine, B. T. \& Li, A., 2007, \apj 657, 810D

\bibitem[Draine \& Lazarian 1998a]{draine1998a} Draine, B. T., \& Lazarian, A. 1998, \apj 494, L19
\bibitem[Draine \& Lazarian 1998b]{draine1998b} Draine, B. T., \& Lazarian, A. 1998 \apj, 508, 157

\bibitem[Dupac 2003]{dupac2003}  Dupac, X.,  Bernard, J.-P, Boudet, N. et al., 2003, \aap, 404, 11D 

\bibitem[Galametz et al. 2009]{galametz2009} Galametz, M., Madden, S.,  Galliano, F. et al.,  2009 \aap  508, 645G

\bibitem[Galliano et al. 2003]{galliano2003}Galliano, F.,  Madden, S. C., Jones, A. P. et al.,  2003, \aap 407, 159G
 
\bibitem[Galliano et al. 2005]{galliano2005}Galliano, F.,  Madden, S. C., Jones, A. P. et al.,  2005, \aap 434, 867G 


\bibitem[Galliano et al. 2008]{galliano2008}Galliano, F., Dwek, E., Chanial, P.,  2008, \apj 672, 214G 

\bibitem[Griffin et al. 2010]{griffin2010} Bendo G. et al., 2010, \aap, this issue 

\bibitem[Gordon 2008]{gordon2008} Gordon, K.D., Engelbracht, C.W., Rieke, G.H. et al., 2008, 2008, \apj 682, 336G
	
\bibitem[Heckman 2001]{heckman2001} Heckman, T. M., Sembach, K. R,; Meurer, G. R. et al., 2001, \apj  554, 1021H

\bibitem[Jones 2009]{jones2009} Jones A.P., \aap 506, 797 

\bibitem[Laor \& Draine 1993]{laor1993} Laor, A. \& Draine, B.T., 1993, \apj 402, 441

\bibitem[Lee \& Skillman 2004]{lee2004} Lee, H. \& Skillman, E.D., 2004, \apj  614, 698L

\bibitem[Lisenfeld et al. 2002]{lisenfeld2002} Lisenfeld, U., Sievers, A., Israel, F., Stil, J., 2002, \aap 382, 860L


\bibitem[Madden 2002]{madden2002}	Madden, Suzanne C., 2002, Ap\&SS, 281, 247M

\bibitem[Mateo 1998]{mateo1998} Mateo, M. 1998, in The Magellanic Clouds and Other Dwarf Galaxies, Proceedings of the Bonn/Bochum-Graduiertenkolleg Workshop, held at the Physikzentrum Bad Honnef, Germany, January 19-22, 1998, Eds.: \ T. Richtler \& J.M. Braun, Shaker Verlag, Aachen, p. 53-66.

\bibitem[Mathis et al. 1993]{mathis2003}  Mathis, J. S., Mezger, P. G., Panagia, N., 1983, \aap 128, 212

\bibitem[Meny et al. 2007]{meny2007} Meny, C.; Gromov, V.; Boudet, N.; Bernard, J.-Ph.; Paradis, D.; Nayral, C., 2007, \aap 468, 171

\bibitem[Meurer et al. 1995]{meurer1995} Meurer, G. R., Heckman, T. M., Leitherer, C. et al., 1995, \aj  110, 2665M

\bibitem[Meurer et al. 1998]{meurer1998} Meurer, G.R., Staveley-Smith, L., Killeen, N. E. B., 1998, \mnras  300, 705M

\bibitem[O'Halloran et al. 2006]{ohalloran2006} O'Halloran B., Satyapal S., Dudik R. P. 2006, \apj 641, 795O

\bibitem[O'Halloran et al. 2008]{ohalloran2008} O'Halloran B., Madden S. C., Abel N. P., 2008, \apj 681, 1205O

\bibitem[Paradis 2009]{paradis2009} Paradis, D., Bernard, J.-Ph.,  M\'eny, C., 2009, \aap 506, 745

\bibitem[Pilbratt 2010]{pilbratt2010} Pilbratt, G. et al., 2010,  \aap, this issue 

\bibitem[Pohlen, 2010]{pohlen2010} Pohlen M., et al., 2010,  \aap, this issue

\bibitem[Poglitsch 2010]{poglitsch2010} Poglitsch, A.. et al., 2010,  \aap, this issue 

\bibitem[Prantzos \& Silk 1998]{prantzos1998} Prantzos, N. \& Silk, J., 1998, \apj  507, 229P

\bibitem[Rouleau \& Martin 1991]{rouleau1991} Rouleau, F. \& Martin, P. G, 1991, \apj 377, 526R

\bibitem[Serra D\'iaz-Cano \& Jones 1998]{serra2008} Serra D\'iaz-Cano, L. \& Jones, A. P., 2008, \aap 492, 127S

\bibitem[Shetty 2009a]{shetty2009a} Shetty, R.,  Kauffmann, J., Schnee, S. \& Goodman, A. A., 2009, \apj 696, 676

\bibitem[Shetty 2009b]{shetty2009b} Shetty, R.,  Kauffmann, J., Schnee, S., Goodman, A. A.,  Ercolano, B., 2009, \apj 696, 2234

\bibitem[Swinyard 2010]{swinyard2010} Swinyard B. et al., 2010, \aap

\bibitem[Tosi et al. 2001]{tosi2001} Tosi, M., Sabbi, E.,  Bellazzini, M. et al., 2001, \aj  122, 1271T

\bibitem[Weingartner \& Draine 2001]{weingartner2001} Weingartner, Joseph C.; Draine, B. T., 2001, \apj 548, 296 

\bibitem[Zubko et al. 2004]{zubko2004}Zubko, V., Dwek, E., Arendt, R.G, 2004, \apjs 152, 211Z

\end{thebibliography}
\end{document}